\documentstyle[preprint,aps,fleqn]{revtex}
\begin{document}
\tolerance 10000

\draft

\title{Spin and Charge Structure Factor of the 2-d Hubbard Model}

\author{Walter Zimmermann$^{1}$, Raymond Fr\'esard$^{2}$ \thanks{Work partially
performed at: Institut f\"ur Theorie der Kondensierten Materie, 
Physikhochhaus, Postfach 6980,
Universit\"at Karlsruhe, 76128 Karlsruhe, Federal Republic of Germany} \thanks
{Present address: Institut de Physique, A-L Breguet 1, CH-2000 Neuch\^atel} 
and Peter 
W\"olfle$^1$}
\vspace*{0.5truecm}

\address{
$^{1}$Institut f\"ur Theorie der Kondensierten Materie, Physikhochhaus, 
Postfach 6980,\\
Universit\"at Karlsruhe, 76128 Karlsruhe, Federal Republic of Germany\\
$^2$ Physics Department, Shimane University, Nishikawatsu-cho 1060,\\
Matsue 690, Shimane, Japan}

\date{Received}
\maketitle
\widetext

\vspace*{-1.0truecm}

\begin{abstract}
\begin{center}
\parbox{14cm}{The spin and charge structure factors are calculated
for the Hubbard model on the square lattice near half-filling using a 
spin-rotation 
invariant six-slave boson representation. The charge structure factor 
shows a broad maximum 
at the zone corner and is found to decrease monotonically with increasing 
interaction 
strength and electron density and increasing temperature. The spin  
structure factor develops with increasing interaction two incommensurate 
peaks at the 
zone boundary and along the zone diagonal. Comparison with results of 
Quantum Monte Carlo and variational calculations is carried out 
and the agreement is found to be good. The limitations of an RPA-type 
approach are pointed 
out.} 
\end{center}
\end{abstract}

\pacs{
\hspace{1.9cm}
 }

%

\section{Introduction}
\noindent

Soon after the discovery of superconductivity in the cuprate materials, it was 
suggested \cite{Anderson} that this phenomenon is closely
related to strong correlation effects. Indeed correlations are responsible for
the insulating state observed in the parent compounds. The simplest
Hamiltonian accounting for such Mott insulators is the one band Hubbard model.
It poses a serious challenge to the theoretician since ordinary
many-body perturbation theory breaks down for strong coupling, being
unable to account for Mott insulator state. A number of new techniques have
been developed, either fully numerical such as Quantum Monte Carlo 
calculations or exact diagonalizations of small systems \cite{Dagotto},
or analytical using the Hubbard X-operator technique
(for a recent work see \cite{Thilo}), the self-consistent 2-particle theory
\cite{Vilk}, the dynamical mean field approximation \cite{Georges} 
or slave bosons. The slave boson method has been applied to a whole range
of problems with local Coulomb interaction: the Kondo impurity model
\cite{Barnes,Coleman}, the Kondo lattice 
model \cite{Coleman,Read,Read/Newns16,klm}, 
the Anderson Hamiltonian
\cite{Barnes,ah} the Hubbard model \cite{Kotliar86,Kotliar} possibly with
orbital degeneracy \cite{RFGK} and even the Bose-Hubbard
model \cite{bhm}. In the Kotliar and 
Ruckenstein (KR) slave boson technique \cite{Kotliar86}
the Gutzwiller Approximation \cite{Gutzwiller63,Brinkman,Vollhardt,Metzner}
appears as a saddle-point 
approximation of this field
theoretical representation of the Hubbard model.
In the latter
a metal-insulator transition occurs at half-filling in the
paramagnetic phase as discussed by Lavagna \cite{Lavagna90}.
The contribution of the thermal fluctuations has been 
calculated \cite{Rasul88} and turned out to be incomplete as this 
representation, even though exact, is not manifestly spin-rotation
invariant. Spin-rotation invariant \cite{Li89} and spin and 
charge-rotation invariant \cite{Fres92a} formulations have been
proposed, all sharing the advantage of treating all the atomic states on an
equal footing, and the first one was used to calculate correlation
functions \cite{Li91} and spin fluctuation contributions to the
specific heat \cite{Woelfle90}. Comparisons of ground state energy
with Quantum Monte-Carlo simulations, including antiferromagnetic
ordering \cite{Lilly} and spiral states \cite{Fresard91}, or with
exact diagonalization data \cite{Fres92b} have been done and yield
excellent agreement, and a magnetic phase diagram has been proposed
\cite{Doll2,Doll1}. The magnetic susceptibility has been evaluated \cite{Doll3}
and shows in the strong coupling regime 
a maximum in its doping dependence in agreement with the 
observed behavior in the cuprates. Such a field theoretical description 
is especially useful
since it allows for the calculation of dynamical quantities as well.
This is the goal of the paper. We first derive expressions for the
spin and charge auto-correlation functions, which we then evaluate
numerically and determine the structure factors. We discuss how they depend
on the doping, the interaction strength and the temperature.

\section{Response functions}

In this work we calculate the spin and the charge structure factors
of the Hubbard model on the square lattice within the Spin Rotation
Invariant (SRI) slave boson formulation \cite{Fres92a}. After having compared
ground state energies, effective band widths and dispersions and
magnetic susceptibilities with results obtained by other techniques,  
generally with very good agreement, structure factors provide another 
test for the approach, which is free of any adjustable parameter.

In the Spin Rotation Invariant slave boson representation, the Hubbard
Hamiltonian is expressed in terms of slave boson 
operators $e_i$, ${\underline p}_i$ $d_i$ for empty, singly occupied
and doubly occupied sites and pseudo-fermions operators $f_{\sigma}$ as
\cite{Kotliar86,Li89,Fres92a}
\begin{equation}
H = \sum_{i,j} t_{i,j} \sum_{\sigma\sigma^{'}\sigma^{''}}
z^{+}_{i\sigma^{''}\sigma}f^{+}_{i\sigma}f_{j\sigma^{'}}
z_{j\sigma^{'}\sigma^{''}} +U\sum_{i}d_{i}^{+}d_{i} 
\end{equation}
where
\begin{equation}
\underline{z}_{i}=e_{i}^{+}\underline{L}_{i}M_{i}
\underline{R}_{i}\ \underline{p}_{i}
+\underline{\tilde p}_{i}^{+}
\underline{L}_{i}M_{i}\underline{R}_{i}d_{i}
\end{equation}
with
\begin{eqnarray}
M_{i}&=&[1+(e_{i}^{+}e_{i}+p_{i0}^{+}p_{i0}+
\vec p_{i}^{+}\cdot \vec p_{i}+d_{i}^{+}d_{i})]^{\frac{1}{2}}
\nonumber\\ 
\underline{L}_{i}&=&[(1-d_{i}^{+}d_{i})\underline{\tau}_0-
\underline{p}_{i}^{+}\underline{p}_{i}]^{-1/2}    
\nonumber\\ 
\underline{R}_{i}&=&[(1-e_{i}^{+}e_{i})\underline{\tau}_0-
\underline{\tilde p}_{i}^{+}\underline{\tilde p}_{i}]^{-1/2}    
\end{eqnarray}
The under-bar denotes a $2\times 2$ matrix in 
spin space, $\underline{\tau}_{0}$ is the unit matrix 
and $\underline{\tilde p}_{i}$ is the time reverse of 
operator $\underline{p}_{i}$.
As usual the slave boson operators have to fulfill constraints. Here they
read
\begin{eqnarray}\label{con1}
e_{i}^{+}e_{i}+(p_{i0}^{+}p_{i0}+
\vec p_{i}^{+} \cdot \vec p_{i})+d_{i}^{+}d_{i}&=&1
\nonumber\\ 
\sum_{\mu}p_{i\mu}^{+}p_{i\mu}+2d_{i}^{+}d_{i}&=&
\sum_{\sigma}f_{i\sigma}^{+}f_{i\sigma}
\nonumber\\ 
(p_{i0}^{+}\vec p_{i}+\vec p_{i}^{+}p_{i0}-
i\vec p_{i}^{+} \times \vec p_{i})&=&
\sum_{\sigma\sigma^{'}}\vec \tau_{\sigma\sigma^{'}}
f_{i\sigma^{'}}^{+}f_{i\sigma}
\end{eqnarray}
and they are respectively enforced by the constraint 
fields $\alpha,\beta_0,\vec \beta$. The scalar and vector slave boson
fields $p_{i0}$ and $\vec p_{i}$ are defined as 
\begin{equation}
\underline p_{i}=\frac{1}{2}(p_{i0}\underline \tau_{0}+
\vec p_{i}\underline{\vec \tau})
\end{equation}
where $\underline{\vec \tau}$ is the vector of Pauli matrices. 
The paramagnetic mean-field free energy results into
\begin{eqnarray}
F& = & -2 T \int d\varepsilon \rho(\varepsilon)\ln[1+\exp-E/T]+ Ud^{2}
\nonumber\\
& & +\alpha(e^{2}+d^{2}+p_{0}^{2}-1)-\beta_{0}(p_{0}+2d^{2})\quad .
\end{eqnarray}
After having fixed the notation
we can proceed to the calculation of the dynamical susceptibilities.
The linear response to an external field is given by the one-loop
order calculation. The expression for the dynamical spin susceptibility
has already been obtained by Li {\it et al} \cite{Li91}, but the calculation of
charge fluctuations turned out to be more involved. In early calculations
it has been assumed that the gauge symmetry group of both KR and SRI
representations allows for gauging away the phases of all slave boson
fields \cite{Rasul88,Lavagna90}. It later turned out that this conclusion 
is erroneous, and one Bose field has to be complex 
\cite{Jolicoeur,Kotliar,Fres92a}. Choosing it to be the field 
describing double occupancy
allows for describing the physics of the upper Hubbard 
band \cite{Castellani}. We note that the fully symmetric gauge approach has
been applied to the calculation of the Landau parameters for 
liquid $^3He$ by Li and B\'enard \cite{Benard}.

The spin and charge auto-correlation functions $\chi_{s}$ and $\chi_{c}$
can be obtained out of:
\begin{equation}
B_{\sigma\sigma^{'}}(i-i^{'},\tau-\tau^{'}) = 
<\hat T [n_{i\sigma}(\tau)n_{i^{'},\sigma^{'}}(\tau^{'})]>
\end{equation}
as
\begin{eqnarray}
\chi_{s}(i-i^{'},\tau-\tau^{'}) & = & \sum_{\sigma\sigma^{'}}
\sigma\sigma^{'}B_{\sigma\sigma^{'}} (i-i^{'},\tau-\tau^{'})\nonumber \\
\chi_{c}(i-i^{'},\tau-\tau^{'})& =& \sum_{\sigma\sigma^{'}} 
B_{\sigma\sigma^{'}}(i-i^{'},\tau-\tau^{'}) \quad
\end{eqnarray}
Using the constraints and the mapping
\begin{equation}
n_{i\sigma}  =  \frac{1}{2}\sum_{\mu=0}^{3}p_{i\mu}^{+}p_{i\mu}+
d_{i}^{+}d_{i} +  
\frac{1}{2}\sigma[p_{i0}^{+}p_{i3}+p_{i3}^{+}p_{i0}-
i(p_{i1}^{+}p_{i2}-p_{i2}^{+}p_{i1})]\quad ,
\end{equation}
one can express the density fluctuations in terms of the slave boson
fields as
\begin{eqnarray}
\sum_{\sigma}\delta n_{\sigma} & = \delta (d^{+} d-e^{+} e) & \equiv \delta N
\nonumber \\
\sum_{\sigma} \sigma \delta n_{\sigma} & = \delta (p_0^{+} p_3 + p_3^{+}p_0)
& \equiv \delta S \quad .
\end{eqnarray}
The correlation functions can be written in terms of the slave boson
correlation functions as:
\begin{eqnarray}\label{korr}
\chi_{s}(k) & = \sum_{\sigma,\sigma^{'}}\sigma\sigma^{'}
<\delta n_{\sigma}(-k)\delta n_{\sigma^{'}}(k)>
& = <\delta S(-k) \delta S(k)>
\nonumber \\
\chi_{c}(k) & = \sum_{\sigma\sigma^{'}}
<\delta n_{\sigma}(-k)\delta n_{\sigma^{'}}(k)>
& = <\delta N (-k) \delta N(k)>\quad.
\end{eqnarray}
Performing the calculation to one-loop order, one can make use of the
propagators given in the appendix to obtain:
\begin{eqnarray}\label{korrend}
\chi_{c}(k) & = & 
2e^{2}S_{11}^{-1}(k)-4edS_{12}^{-1}(k)+2d^{2}S_{22}^{-1}(k) 
\nonumber \\
\chi_{s}(k) & = & 2p_{0}^{2}S_{77}^{-1}(k)\quad .
\end{eqnarray}
Including the inverse matrix elements, we get 
\begin{equation}
\chi_{c}(k)=\frac{
\bigg(
S_{33}(k)S_{55}(k)e^{2}[-2p_{0}^{2}\Gamma_{1}(k)+
8dp_{0}\Gamma_{2}(k)-8d^{2}\Gamma_{3}(k)]-
2e^{4}p_{0}^{2}S_{55}^{2}(k)\omega^{2}\bigg)}
{S_{33}(k)(\Gamma_{1}(k)\Gamma_{3}(k)-\Gamma_{2}^{2}(k))+
\omega^{2}\Gamma_{3}(k)S_{55}(k)e^{2} }
\end{equation}
and
\begin{equation} \label{chis}
\chi_{s}(k)=\frac{\chi_{0}(k)}
{1+A_{\vec k}\chi_{0}(k)+A_{1}\chi_{1}(k)+
A_{2}[\chi_{1}^{2}(k)-\chi_{0}(k)\chi_{2}(k)]}\quad ,
\end{equation}
where
\begin{eqnarray}\label{theas}
A_{k} & = & (2p_{0}^{2})^{-1}\bigg[
\alpha-\beta_{0}+\varepsilon_{0}z_{0}
\frac{\partial^{2}z_{\uparrow}}{\partial p_{3}^{2}}+
\varepsilon_{\vec k}(\frac{\partial z_{\uparrow}}
{\partial p_{3}})^{2}\bigg]
\nonumber \\
A_{1} & = & p_{0}^{-1}z_{0}\frac{\partial 
z_{\uparrow}}{\partial p_{3}}
\nonumber \\
A_{2} & = & (4p_{0}^{2})^{-1}z_{0}^{2}(\frac{\partial z_{\uparrow}}
{\partial p_{3}})^{2}\quad ,
\end{eqnarray}
and
\begin{eqnarray}\label{gammas}
\Gamma_{1}(k) & = & -S_{55}(k)(e^{2}S_{22}(k)-2edS_{12}(k)+d^{2}S_{11}(k))+
(eS_{25}(k)-dS_{15}(k))^{2}
\nonumber \\
\Gamma_{2}(k) & = & -S_{55}(k)(e^{2}S_{24}(k)-p_{0}eS_{12}(k)-edS_{14}(k)
+dp_{0}S_{11}(k))\nonumber \\
&+& (eS_{25}(k)-dS_{15}(k))(eS_{45}(k)-p_{0}S_{15}(k))
\nonumber \\
\Gamma_{3}(k) & = & 
-S_{55}(k)(e^{2}S_{44}(k)-2ep_{0}S_{14}(k)+p_{0}^{2}S_{11}(k))+
(eS_{45}(k)-p_{0}S_{15}(k))^{2} \quad .
\end{eqnarray}

In the following we shall evaluate numerically the spin and charge structure
factors:
\begin{equation} \label{spinsf}
S_{x}({\vec q}) = -\int\limits_{-\infty}^{+\infty}\frac{d\omega}{\pi}
\frac{Im \chi_{x}({\vec q},\omega+i0)}
{1-exp(-\omega/T)} ; \qquad x=s,c \quad .
\end{equation}
Note that while deriving the expressions (\ref{korrend}) for the
spin and charge correlation functions, we are dealing with both complex
($d$) and real ($e,p_{\mu}$) fields. For the latter, following an argument
by Read and Newns \cite{Read/Newns16}, the contribution from the
measure to the action ($\sum_{i,n}\ln{(e_{i,n} \prod_{\mu} p^{\mu}_{i,n})}$)
has been neglected. This aspect has been recently re-investigated in
the framework of the $1/N$ expansion of the large $U$ Hubbard model
\cite{LGRF}. It has been shown that including this contribution only
leads to a minor change of the action, leaving its numerical
value unchanged. This argument holds in the present context as well.
\section{Results}
We first determine the spin structure factor as given by Eq. (\ref{spinsf}).
In the weak coupling limit our result reduces to the RPA \cite{LiPRB}, and
thus for small coupling and particle densities, the agreement with the
exact solution is expected to be very good. However the RPA is getting
less and less reliable as the interaction strength increases. This leads
to unreasonable results in RPA as e.g. a magnetic instability of the
paramagnetic state for any particle filling above a critical coupling
\cite{Dzierzawa}. Such a deficiency is corrected in slave boson mean field
theory. At zero temperature magnetic instabilities appear only beyond
a certain particle filling \cite{Doll1}. For increasing temperature the
region of magnetic long-range order is shrinking rapidly
\cite{Doll3}. Strictly speaking, long-range order is absent at any
finite temperature due to thermal excitations of spin waves
(Mermin-Wagner theorem), which are not taken into account in
slave boson mean-field theory.

Here we display results in the paramagnetic state, away from the
instability line. Let us start by comparing (Fig. 1) our result for the 
spin structure factor to Quantum Monte
Carlo results \cite{Michael}. Here the structure factor is plotted as a 
function 
of wave number along straight lines $\Gamma$-$X$-$M$-$\Gamma$ in the 
Brillouin zone.  
For $U=4t$, temperature $T=t/6$ ($\beta=6$; $\beta=t/T$) and a doping 
$\delta = 0.275$,
the overall agreement is good and the trends of the simulations are reproduced.
Especially the
peak is very broad and is centered around $(3\pi/4,\pi)$ for the QMC data. 
Out of our approach we get 
two peaks located away from $(\pi,\pi)$, which are indicating the onset of 
incommensurate short 
ranged spin order. As compared to the QMC calculations we obtain a second peak 
along the diagonal of the Brillouin zone which cannot be resolved in the 
simulations due to the small size of the system. The height of the peak which 
is located along the zone boundary is larger than the one of the peak which is
lying on the diagonal of the Brillouin zone. We note that we are comparing 
the slave boson results for an infinite system to the raw QMC data on the
$8\times8$ lattice. We believe that this comparison is meaningful since we 
observed that finite size effects are small (in the percent range as compared
to the $6\times6$ lattice) for the set of parameters we are using. The 
agreement between the 2 approaches is mostly qualitative, and the slave boson 
calculation tends to overestimate the tendency towards magnetic ordering. But 
it also allows for gaining additional informations which are not revealed by 
the simulations like the presence of a second peak along the diagonal of the
Brillouin zone.
QMC data are not available in the complete parameter 
range. As an alternative there exist variational approximations at $T=0$ 
to which we can compare our data.
A very promising variational ansatz is provided by a generalized 
Baeriswil-Gutzwiller wave-function \cite{Otsuka}. In Fig. 2 we compare our 
result for $U=2t$ and $U=4t$ ($\beta=8$, because of onset of long
range order in our mean-field treatment) 
and $\delta=0.218$ with the zero temperature results
 of ref.\cite{Otsuka}, and 
we reach the same conclusions as when comparing with the QMC simulations. On
top of that we notice that the agreement is better for weaker interaction.
Note that the position of the peaks
is temperature dependent. At low temperature it is systematically
located away from $(\pi,\pi)$, except at half-filling, but it moves 
towards $(\pi,\pi)$ for increasing temperature, 
as shown on Fig. 3 for $U=4t$. The first result
of a small increase of the temperature is to suppress the value of the
structure factor at its peak position, while the one at the zone
corner is increasing, up to the point where it becomes the dominant
one, as emphasized in the inset of Fig. 3. Increasing the temperature further 
results into an overall reduction
of the structure factor for large momenta, and an overall increase of it for
small momenta. The fluctuation-dissipation theorem relates the $q=0$ value
of the spin structure factor to the magnetic susceptibility. As shown in
the inset of Fig. 3, the latter is growing with increasing temperature for
low $T$, reaches a maximum at $\beta \sim 4$, and decreases beyond. This
behavior is reminiscent of the spin gap behavior observed in the High $T_c$
superconductors, although there it occurs at a smaller energy scale.
As one can see from Fig. 3 , the spin structure factor appears to approach 
its zero temperature limiting value at $\beta \simeq 8$. We may therefore use 
the result for 
$U=4t$ at $\beta=8$ instead of the $T=0$ ($\beta = \infty$) result (which is 
not 
accessible because of onset of long range order in our mean-field treatment), 
to compare with the variational Monte Carlo (VMC) result at doping 
concentration $\delta=0.218$ (Fig. 2). Whether incommensurate long range
order occurs off half-filling using Otsuka's wave-function is not known. In
any case it sets in using the Gutzwiller wave-function \cite{Giamarchi}.

We now consider the dependence of the spin structure factor on the 
interaction and the density. Here we fix the temperature to $\beta=8$, and
we calculate $S_s(q)$ for $U=t$ and $U=2t$ and display it on Fig. 4 for 
$\delta =0.1$, on Fig. 5 for $\delta =0.2$ and $U/t =$ 1, 2, 3 and 4, 
and on Fig. 6 for $\delta =0.3$ and $U/t =$ 2, 4, 6, 7 and 8.
In all cases we obtain that raising up the interaction generates more spin
ordering, and the structures in the curves are much more pronounced. 
At this temperature the position of the peak remains at the zone corner for 
small
doping (see Fig. 4), while it is 
shifted away from its commensurate value  for larger doping (see Fig. 5 and 6).
The influence of the 
interaction is stronger
when the system is denser, and this effect is enhanced by the vicinity of the
perfect nesting point. For large doping the peaks become much 
broader resulting in an incommensurate very short ranged spin order. Also
increasing the hole doping shifts the positions of the maxima of $S_s(q)$
further away from $(\pi,\pi)$, both  along the $M-X$ line and along the 
$\Gamma-M$ line.
One may ask to what extent the results presented above can be obtained in the 
framework of an RPA-type (or Fermi liquid type) scheme. For this purpose we 
compare the result for $\chi_{s}$ obtained by replacing the effective 
interaction 
$A_{\vec k}({\vec q})$ by its $q=0$ limit and putting $A_{1}$ and $A_{2}$ 
equal to 
zero in Eq. (\ref{chis},\ref{theas}). This allows for 
investigating the influence of the 
momentum 
dependence of the effective interaction on the spin order. The result is 
displayed on Fig. 7 for $U/t=2, 4$ and $6$. Clearly the approximation is 
very good for moderate couplings, but gets gradually worse for increasing 
interaction. Thus the dispersion of the effective interaction has a negligible
influence on the structure factor for moderate interaction, but an important
one for intermediate to large couplings, where it strongly shifts and 
suppresses the peaks of the structure factor. We thus conclude that knowing
the Landau parameter $F_0^a$ is not all what is 
needed in order to determine the spin structure 
factor, especially when the fluctuations have a dominant short wavelength
character.

We now turn to the charge structure factor. In 
Fig. 8 we compare our result for $U=4t$ and $\delta=0.275$ at temperature 
$T=t/6$ with 
the Quantum Monte Carlo result of Dzierzawa \cite{Michael}. The charge 
structure factor consists of one broad peak which is centered at $(\pi,\pi)$.
Fig. 8 clearly shows that the agreement between both approaches is excellent,
and that the difference does not exceed a few percent. 
In Fig. 9 our result at zero temperature and $\delta=0.218$ for $U=4t$, 
$U=8t$, and $U=16t$ are compared with the variational Monte Carlo result of 
\cite{Otsuka}, and the dependence on U is also displayed on Fig. 10. 
As 
expected, increasing the interaction strength U leads to a suppression 
and a further broadening of the charge structure factor. 
In Fig. 11, the 
dependence on doping is shown for $U=8t$ at zero temperature. In weak 
coupling, one would expect $\chi_{c}$ to decrease upon doping. However
the opposite behavior holds in a dense strongly correlated system, and our
approach succeeds in obtaining this subtle effect. As compared to 
the 2-particle 
self-consistent theory \cite{Vilk}, the reason of the success is however
quite different. Indeed if we use the ``RPA'' approximation (the same 
procedure as above, but now in the charge
fluctuation sector), we again obtain that the difference as compared to the
full expression grows under an increase in the interaction strength as
displayed in Fig. 12. Thus in our theory it is essential to take the dispersion
of the interaction into account in order to obtain a good result, while
such a dispersion is neglected in the 2-particle 
self-consistent theory \cite{Vilk}.
We also performed the calculation at finite
temperature. In contrast to the spin structure factor, the charge structure 
factor is
mostly temperature independent at low $T$. This is a pure interaction
effect, since in the non-interacting limit both structure factors have the
same $T$-dependence.
A further increase in $T$
simply leads to an overall reduction of the charge structure factor for large
momenta, and an increase of it for small momenta, as shown
in Fig. 13, where $S_{c}(q)$ is plotted for $U=4t$ and $\delta=0.275$ at 
temperatures ranging 
from $T=t/1.3$ down to $T=t/8$. Furthermore the temperature has no influence
on the position of the peak of the charge structure factor. The 
fluctuation-dissipation theorem relates the charge structure factor at $q=0$
to the charge susceptibility. As indicated in the inset of Fig. 13, the
latter decreases monotonously with increasing Temperature, in agreement
with Fermi liquid theory.

%
\section{Summary}
In this work we derived and evaluated the spin and charge structure
factors of the Hubbard Model within the Spin Rotation Invariant 
six-slave boson formulation
of the Hubbard model. We considered Gaussian fluctuations about the
paramagnetic saddle-point, at small hole doping and for finite
temperature, where the mean field solution is the paramagnetic
one. The agreement with available exact numerical results for finite 
size systems and for variational wave functions was found to be 
very good. It is found that increasing either the interaction strength or the
density leads to more incommensurate spin order, and less commensurate charge
order. 
The slave boson mean field theory appears to provide a good 
starting point for describing dynamical correlations on the level of 
Gaussian fluctuations. It would be of interest to extend the calculations 
to the
magnetically ordered phases, discussed e.g. in
ref. \cite{Fres92b}. Also the consideration of short range
antiferromagnetic order \cite{Trapper} is presumably important at low 
temperature, and should be taken into account.
  
\section{Acknowledgments}
This work has been supported by the Deutsche Forschungsgemeinschaft 
through Sonderforschungsbereich 195. One of
us (RF) is grateful to the Fonds National Suisse de la Recherche Scientifique
for financial support under Grant 8220-0284525 and to the Japanese Ministry
of Education for a Grant-in-Aid for scientific research No. A 07740301.

\section{Appendix}
In second order in the bosonic variables the action is given by \cite{Fres92a}:
\begin{equation}
S = \sum_{q,\mu,\nu}  \psi_{\mu}(-q)S_{\mu,\nu}(q)\psi_{\nu}(q)
\end{equation}
with $\psi_{1}=e, \psi_{2}=d', \psi_{3}= d'', \psi_{4}= p_0, 
\psi_{5} = \beta_0, \psi_{6} = \alpha, \psi_{7} = p_1, \psi_{8} = \beta_1,
\psi_{9} = p_2, \psi_{10} = \beta_2, \psi_{11} = p_3, \psi_{12} = \beta_3$.
Here $d'$ and $d''$ are the real and imaginary parts of the complex 
d-field. The propagator matrix decouples
into 4 blocks, one for the charge fluctuations and 3 for the spin
fluctuations. Since the calculation is straightforward and most results
can be gathered from ref \cite{Li91}, we only quote the results.
The charge part of the fluctuation matrix is given by:
\begin{eqnarray}
S_{11}(k) & = & \alpha+\tilde S_{11}(k)
\nonumber \\
S_{22}(k) & = & \alpha-2\beta_{0}+U+\tilde S_{22}(k)
\nonumber \\
S_{33}(k) & = & \alpha-2\beta_{0}+U+\tilde S_{33}^{'}(k)
\nonumber \\
S_{23}(k) & = & \omega_{n}
\nonumber \\
S_{44}(k) & = & \alpha-\beta_{0}+\tilde S_{44}^{'}(k)
\nonumber \\
S_{ij}(k) & = & \tilde S_{ij}(k);i\not= j;i,j=1,2,4
\nonumber \\
S_{15}(k) & = & -\frac{1}{2} \chi_{1}(k)z_{0}
\frac{\partial z}{\partial e}
\nonumber \\
S_{16}(k) & = & e
\nonumber \\
S_{25}(k) & = & -2d-\frac{1}{2}\chi_{1}(k) 
\frac{\partial z}{\partial d'}z_{0} 
\nonumber \\
S_{26}(k) & = & d
\nonumber \\
S_{45}(k) & = & -p_{0}-\frac{1}{2}\chi_{1}(k)
\frac{\partial z}{\partial p_{0}}z_{0}
\nonumber \\
S_{46}(k) & = & p_{0}
\nonumber \\
S_{55}(k) & = & -\frac{1}{2}\chi_{0}(k)
\end{eqnarray}
and the spin part by:
\begin{eqnarray}
S_{77}(k) & = & S_{99}(k)=S_{11.11}(k)=
\alpha-\beta_{0}+\tilde S_{77}(k)
\nonumber \\
S_{88}(k) & = & S_{10,10}(k)=S_{12,12}(k)=
-\frac{1}{2}\chi_{0}(k)
\nonumber \\
S_{78}(k) & = & S_{9,10}(k)=S_{11,12}(k)=
-p_{0}-\frac{1}{2}\chi_{1}(k)
\frac{\partial z_{\uparrow}}{\partial p_{3}}z_{0} \quad .
\end{eqnarray}
We also defined:
\begin{eqnarray}\label{sti}
\tilde S_{\mu\nu}(k) & = & \varepsilon_{0} z_{0}\frac{\partial^{2} z}
{\partial \psi_{\mu} \partial \psi_{\nu}}+
[\varepsilon_{\vec k}-\frac{1}{2}z_{0}^{2}\chi_{2}(k)]
\frac{\partial z}{\partial \psi_{\mu}}
\frac{\partial z}{\partial \psi_{\nu}} \quad \mu, \nu=1,2,4,7,9,11
\nonumber \\
\tilde S_{33}^{'}(k) & = & \varepsilon_{0} z_{0}\frac{\partial^{2} z}
{\partial d'' \partial d''}+
[\varepsilon_{\vec k}-\frac{1}{2}z_{0}^{2}\chi_{2}^{'}(k)]
\frac{\partial z^{+}}{\partial d^{I}}
\frac{\partial z}{\partial d^{I}}
\end{eqnarray}
and
\begin{equation}\label{epsdede}
\varepsilon_{\vec k}=\sum_{p \sigma}t_{\vec p - \vec k} G_{0}(p)
\end{equation}
in terms of the pseudo-fermion Greens function $G_{0}(p)
=1/(i\omega_{n}-E_{\vec k})$ and the effective dispersion:
\begin{equation}
E_{\vec k}=z_{0}^{2}t_{\vec k}-\mu+\beta_{0}\quad .
\end{equation}
We also introduced the dynamical response functions of the fermionic
system:
\begin{eqnarray}\label{dynrf}
\chi_{n}(k) & = & -\sum_{p \sigma}(t_{\vec p}+t_{\vec p +\vec k})^{n}
G_{0\sigma}(p)G_{0\sigma}(p+k)=\chi_{n}(-k) (n=0,1,2)
\nonumber \\
\chi_{2}^{'}(k) & = & -\sum_{p\sigma}
(t_{\vec p+ \vec k}-t_{\vec p })^{2}
G_{0\sigma}(p)G_{0\sigma}(p+k)\quad .
\end{eqnarray}
In Eq. (\ref{sti}) the derivatives are given by:
\begin{eqnarray}
\frac{\partial z}{\partial d'}& = &\sqrt{2}p_0 \eta 
(1 + \frac{2xd}{1+\delta}) \nonumber \\
\frac{\partial z}{\partial d''}& = &i\sqrt{2}p_0 \eta \nonumber \\
\frac{\partial^{2} z}{\partial d'^{2}} & = &
\frac{2\sqrt{2}p_0 \eta}{1+\delta} (2d + x + \frac{3xd^2}{1+\delta})
\nonumber \\
\frac{\partial^2 z}{\partial d' \partial d''}& = & i
\frac{ 2\sqrt{2}p_0 d\eta}{1+\delta} \nonumber \\
\frac{\partial^{2} z}{\partial d''^{2}} & = &
\frac{ 2\sqrt{2}p_0 x\eta}{1+\delta} \nonumber \\
\frac{\partial^2 z}{\partial d' \partial e}& = & 
2\sqrt{2}p_0 \eta( \frac{e}{1-\delta}+ \frac{d}{1+\delta} + 2edx\eta^2)
\nonumber \\
\frac{\partial^2 z}{\partial d' \partial p_0}& = & 
\sqrt{2} \eta( 1 + 2 p_0^2 \eta^2 + \frac{2xd}{1+\delta}+ 
\frac{6p_0^2xd}{(1+\delta)^2}+2p_0^2xd\eta^2)\nonumber \\
\frac{\partial^2 z}{\partial d'' \partial e}& = & i
\frac{ 2\sqrt{2}p_0 e\eta}{1-\delta} \nonumber \\
\frac{\partial^2 z}{\partial d'' \partial p_0}& = & i
\sqrt{2} \eta( 1 + 2 p_0^2 \eta^2 )
\end{eqnarray}
With $x=e+d$ and $\eta^{2}=1/(1-\delta^{2})$. The other derivatives
can be found in Ref. \cite{Li91}. This extends the result of Bang {\it et al}
\cite{Kotliar} to the metallic regime.

\newpage
{\bf FIGURE CAPTIONS}
\vspace{3cm}

FIG. 1. Comparison of the Quantum Monte Carlo (circles) and Slave Boson
(full line) spin structure factors for $U=4t$, $\delta=0.275$ and $\beta=6$.

FIG. 2. Comparison of the Variational Monte Carlo (circles for $U=2t$ and 
squares for $U=4t$ ) and Slave Boson (dashed line for $U=2t$, $\beta=8$ and 
full line for 
$U=4t$, $\beta=8$) 
spin structure factors for $\delta=0.218$ and $T=0$.

FIG.3. Temperature dependence of the spin structure factor for $U=4t$, 
and $\delta=0.275$. The temperatures are $\beta=$ 8 (full line), 4 (dotted line), 2 (dashed-dotted line) and 1.3 (long dashed- short 
dashed line). Left inset: Magnification of the latter around the M-point.
Right inset: Temperature dependence of the magnetic susceptibility for $U=4t$
and $\delta=0.275$.

FIG. 4. U-dependence of the Slave Boson spin structure factor for $U=1t$ 
(circles) and $U=2t$ (triangles),  for $\delta=0.1$ and $\beta=8$.

FIG. 5. U-dependence of the Slave Boson spin structure factor for $U=1t$ 
(circles), $U=2t$ (triangles), $U=3t$ (pluses) and $U=4t$ (x's), 
for $\delta=0.2$ and $\beta=8$.

FIG. 6. U-dependence of the Slave Boson spin structure factor for $U=2t$ 
(circles), $U=4t$ (triangles), $U=6t$ (pluses), $U=7t$ (x's) and 
$U=8t$ (diamonds), for $\delta=0.3$ and 
$\beta=8$.

FIG. 7. Comparison of the Slave Boson (x's resp. diamonds resp. stars) and RPA 
(circles resp. triangles resp. pluses) spin structure factors for
$U=2t$ resp. $U=4t$ resp.  $U=6t$, and
$\delta=0.3$ and $\beta=8$.

FIG. 8. Comparison of the Quantum Monte Carlo (triangles) and Slave Boson
(full line) charge structure factors for $U=4t$, $\delta=0.275$ and $\beta=6$.

FIG. 9. Comparison of the Variational Monte Carlo (circles, pluses, diamonds) 
and Slave Boson(full, dashed, dashed-dotted line) charge structure factors for 
$U=4t$, $U=8t$ and $U=16t$, 
and $\delta=0.218$ and $T=0$.

FIG. 10. U-dependence of the Slave Boson charge structure factor for $U=2t$ 
(circles), $U=4t$ (triangles), $U=8t$ (pluses) and $U=16t$ (x's),
for $\delta=0.275$ and $T=0$.

FIG. 11. Doping-dependence of the Slave Boson charge structure factor for
$\delta=0.1$ (circles), $\delta=0.2$ (triangles), and $\delta=0.3$ (pluses),
and for $U=8t$ and $T=0$.

FIG. 12. Comparison of the Slave Boson (circles, pluses, diamonds)
and RPA (squares, triangles, x's) charge structure factors 
for $U=2t$ resp. $U=4t$, resp. $U=8t$, and $\delta=0.218$ and $T=0$.

FIG. 13. Temperature dependence of the charge structure factor for $U=4t$, 
and $\delta=0.275$. The temperatures are $\beta=$ 8 (circles), 6 (triangles),
4 (pluses), 2 (x's) and 1.3 (diamonds). Inset: Temperature dependence of
the charge susceptibility  for $U=4t$ and $\delta=0.275$.
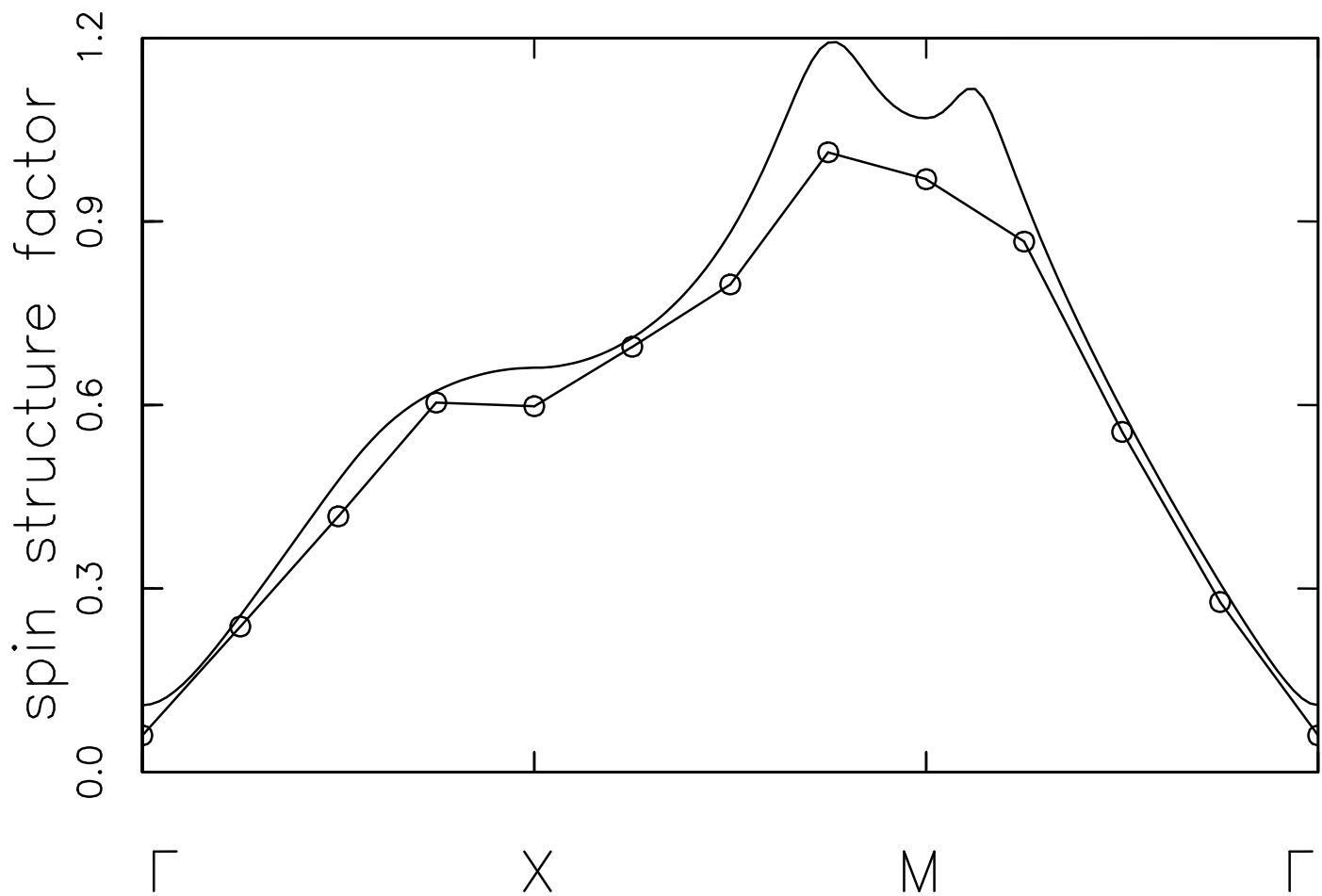
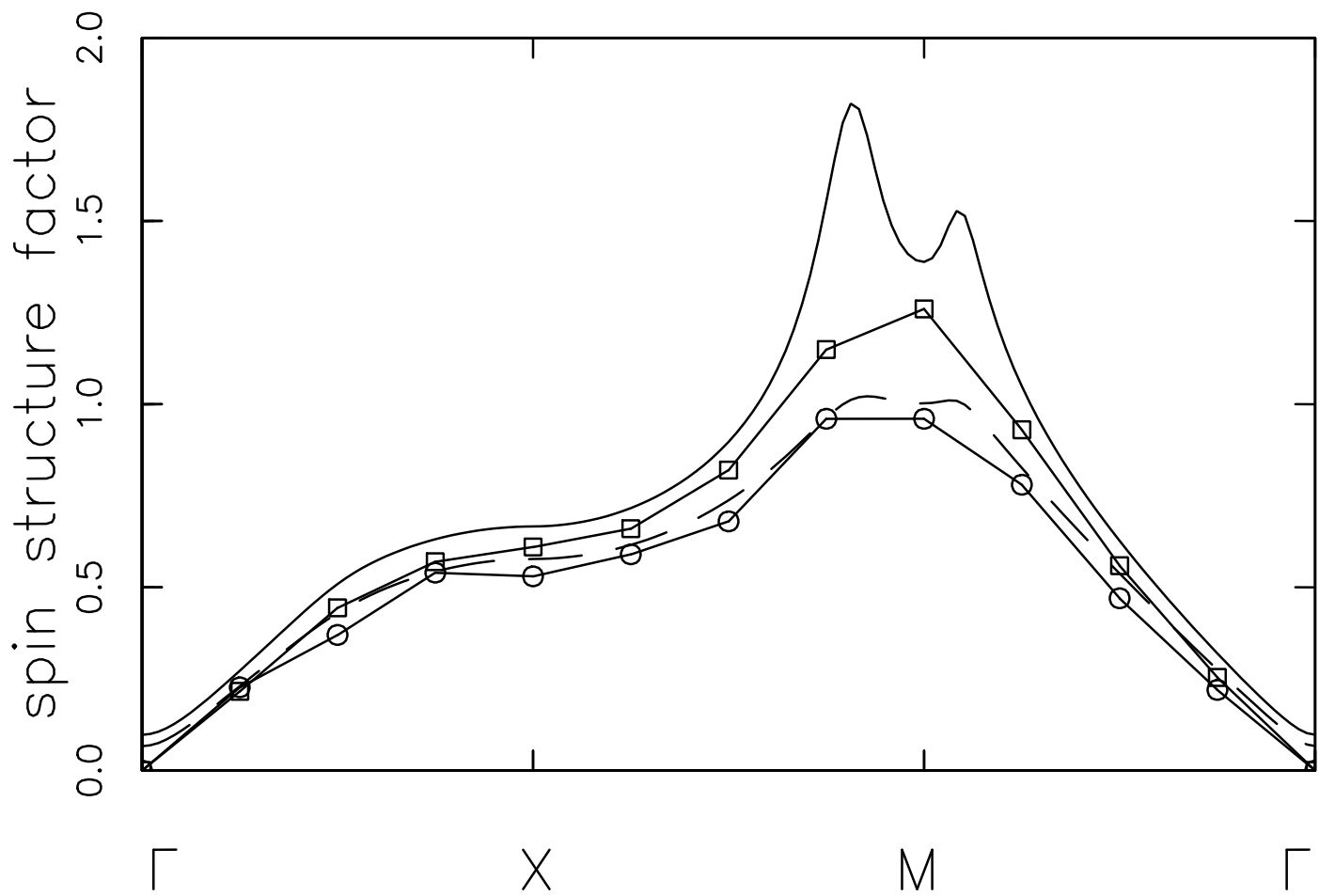
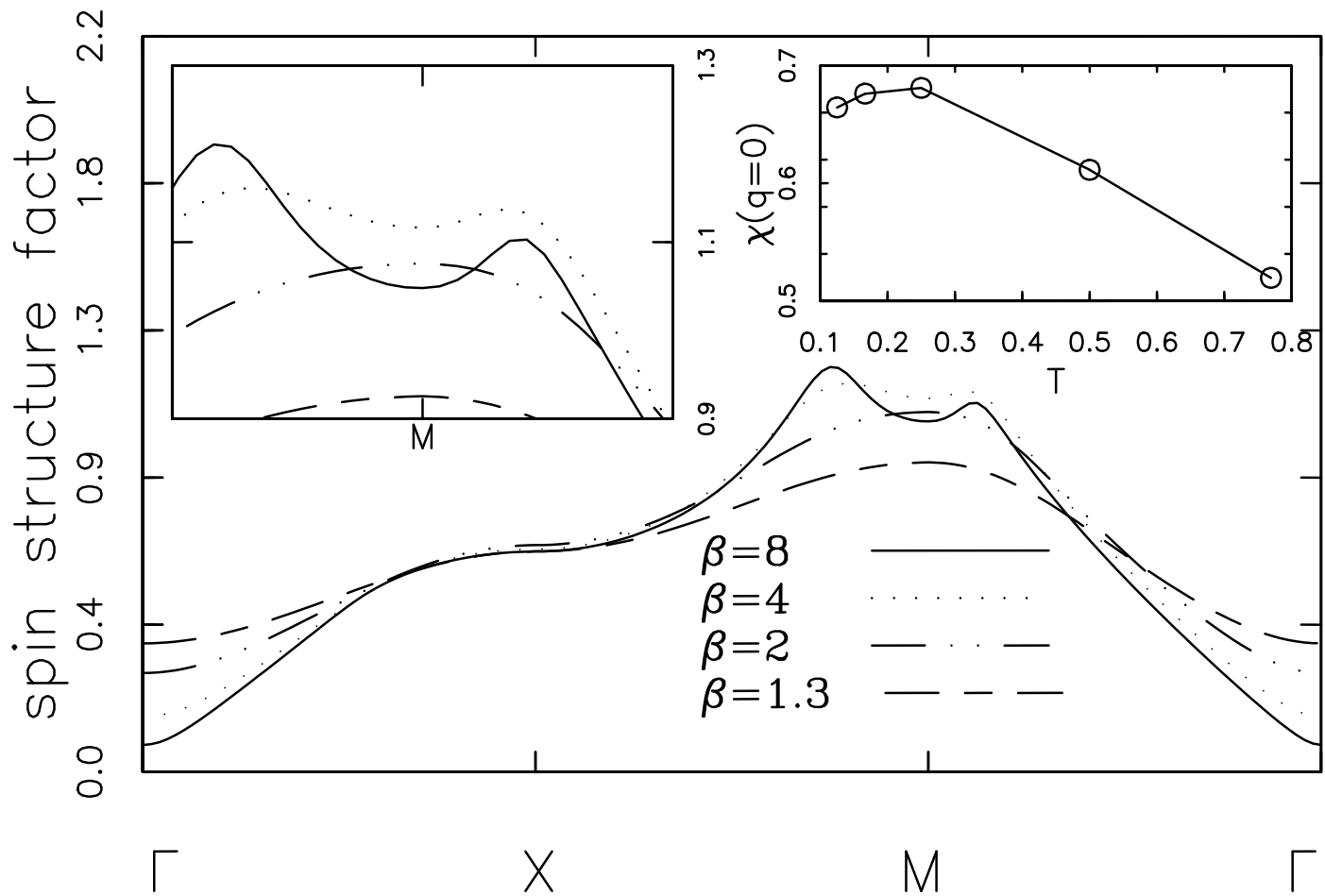
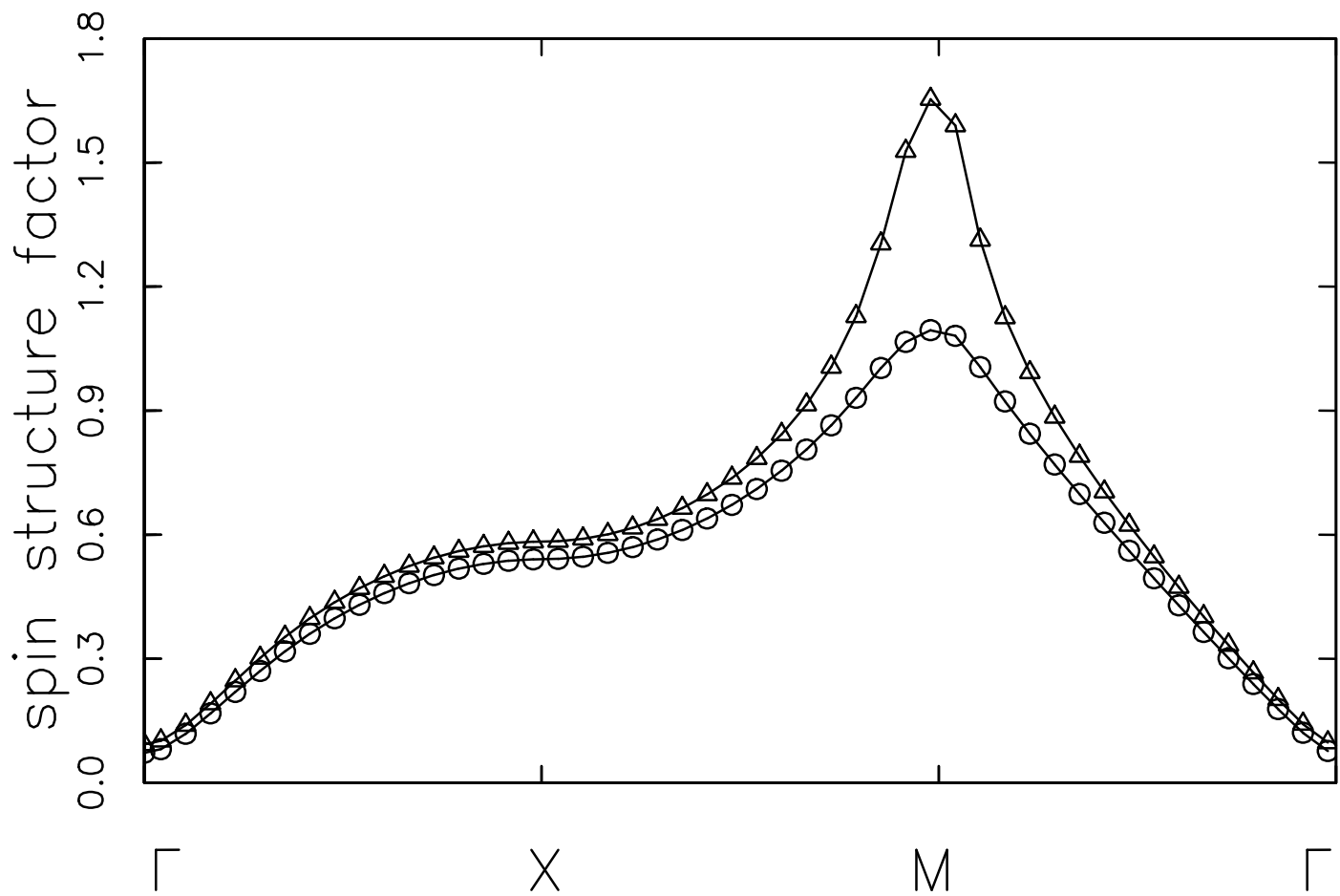
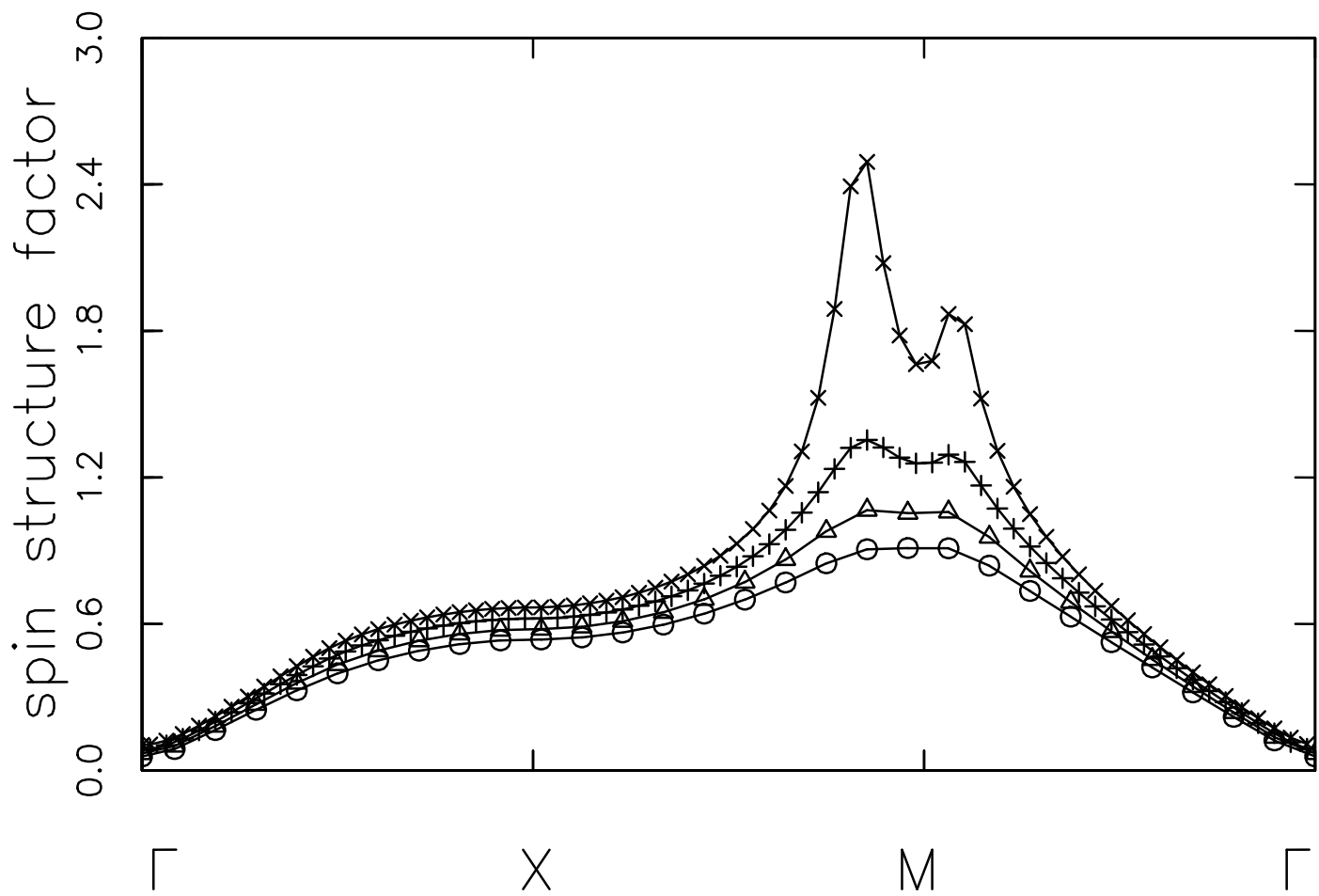
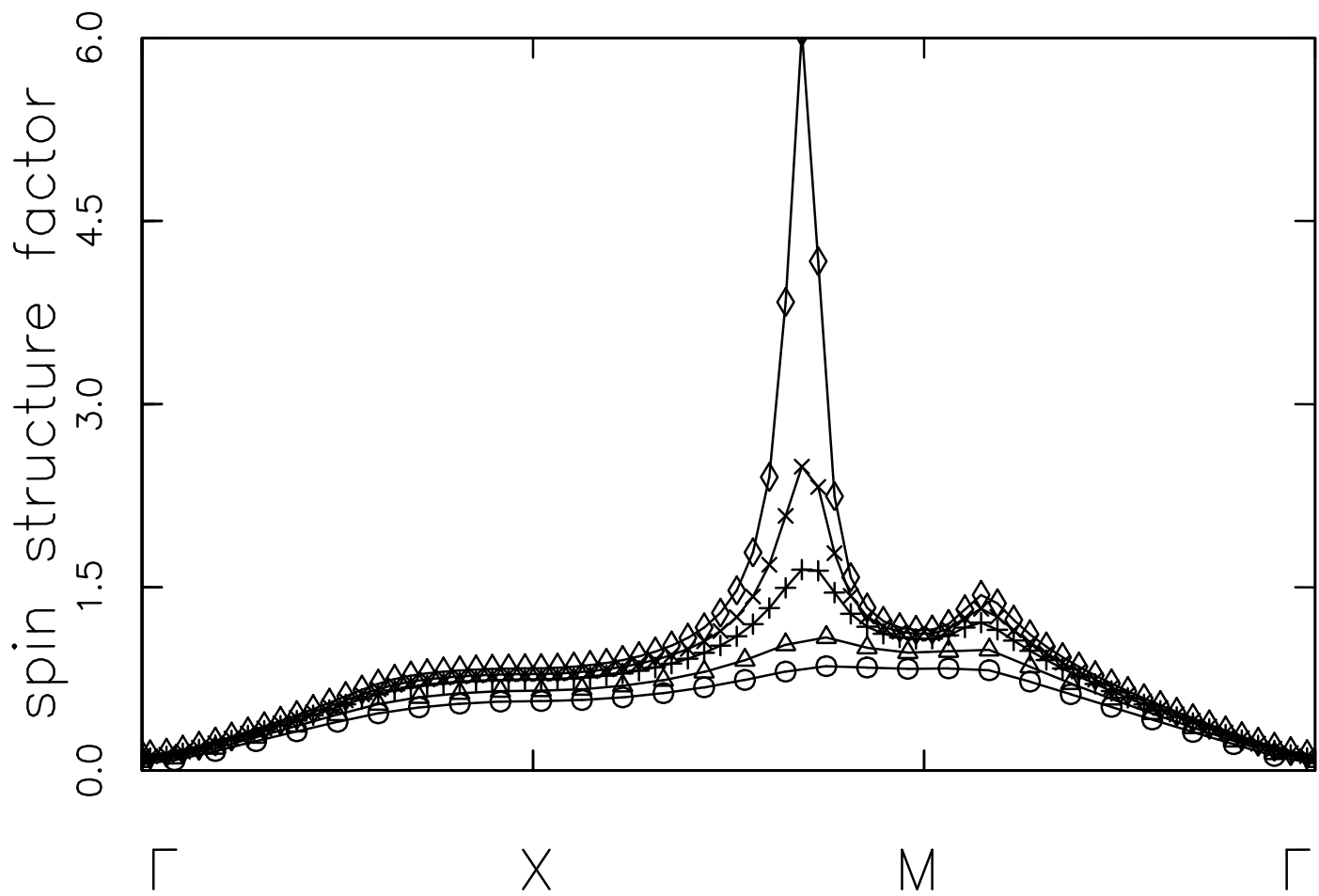
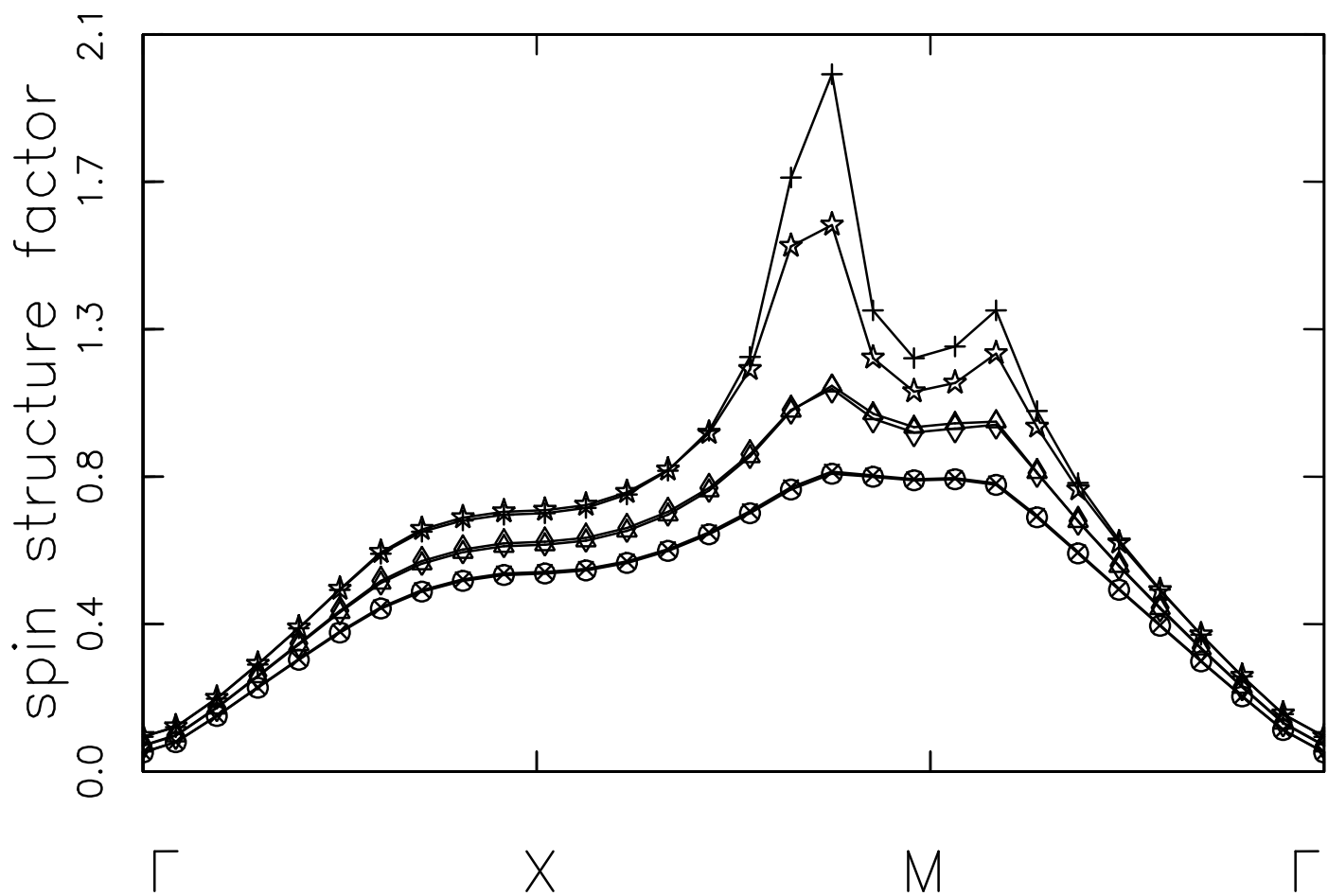
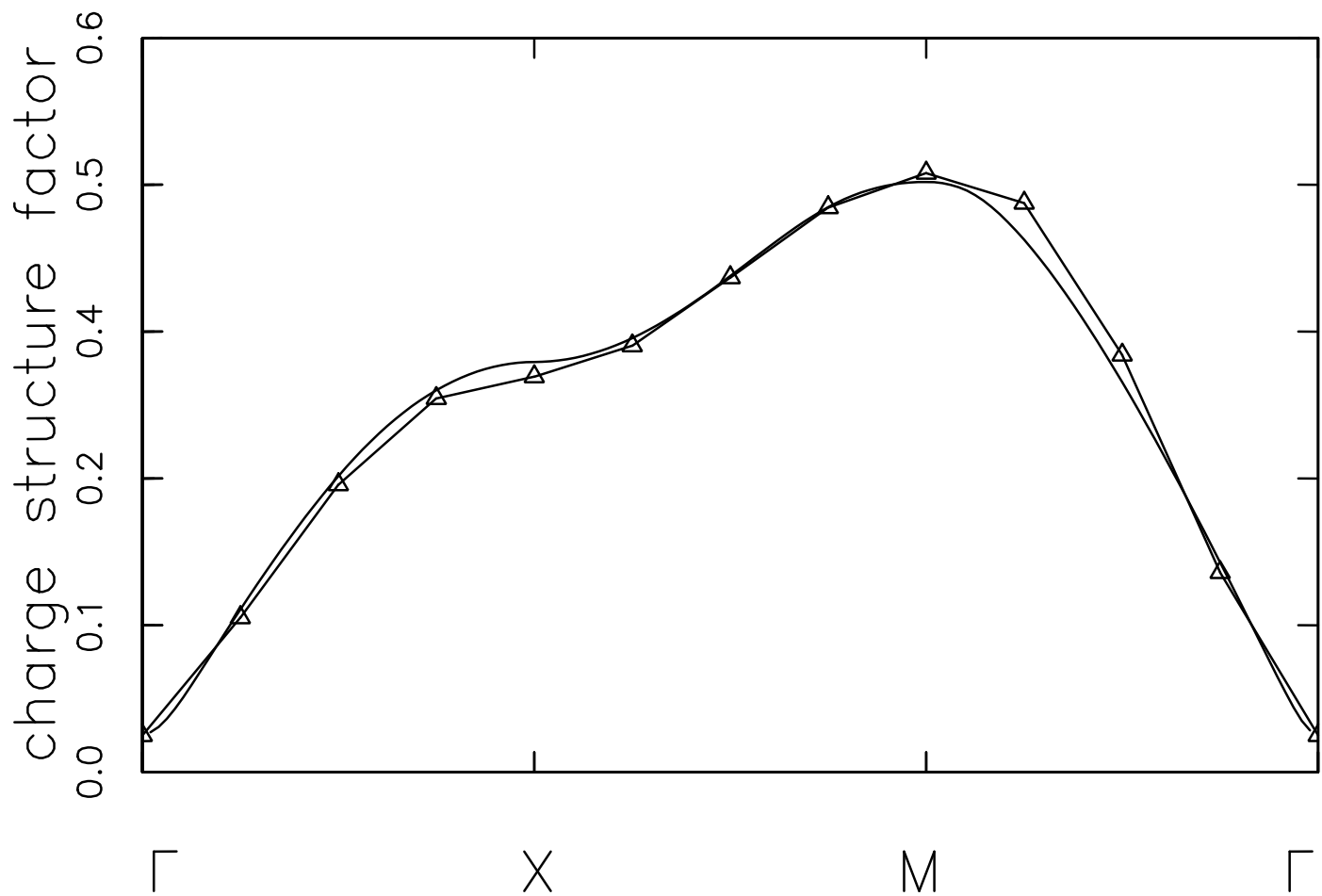
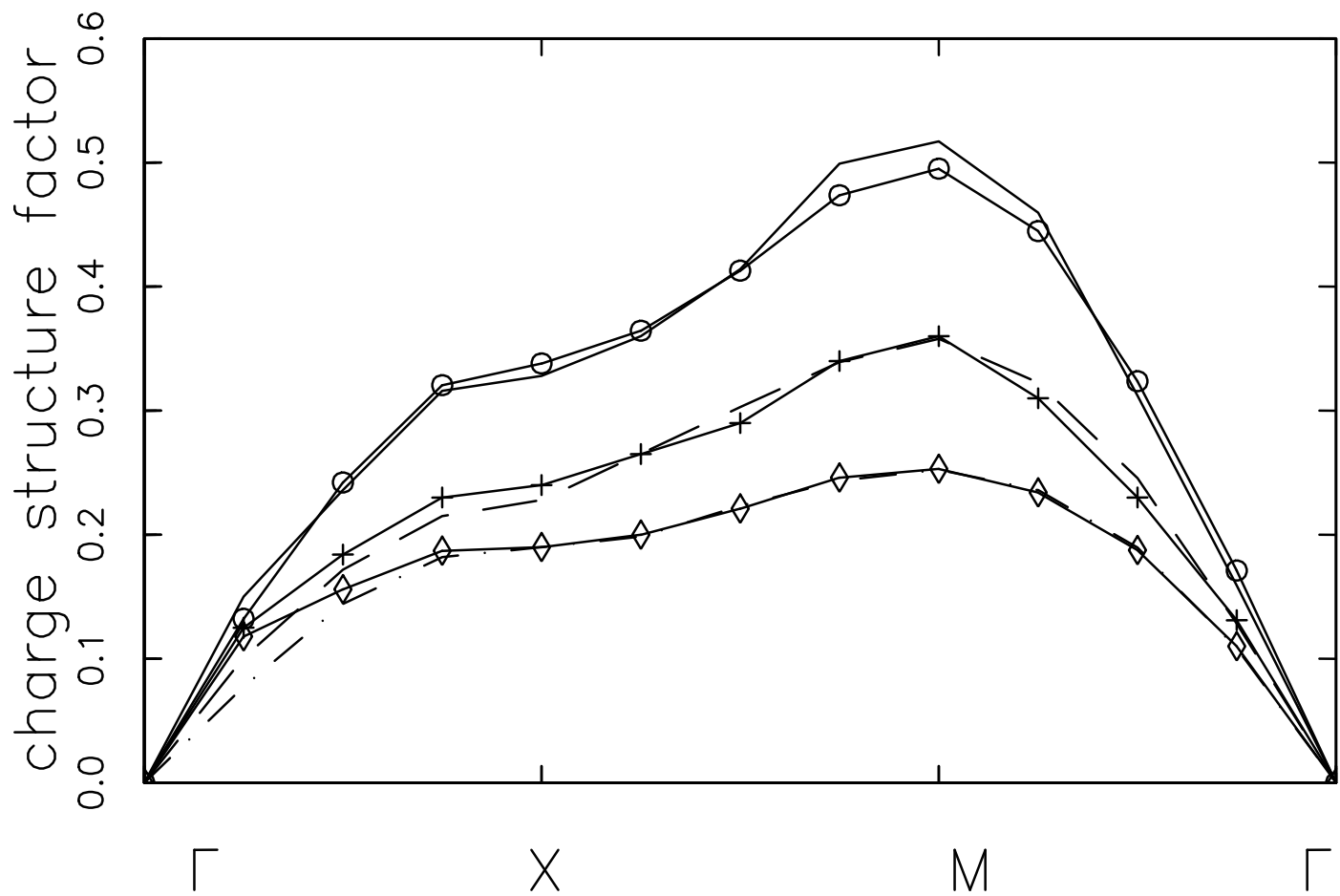
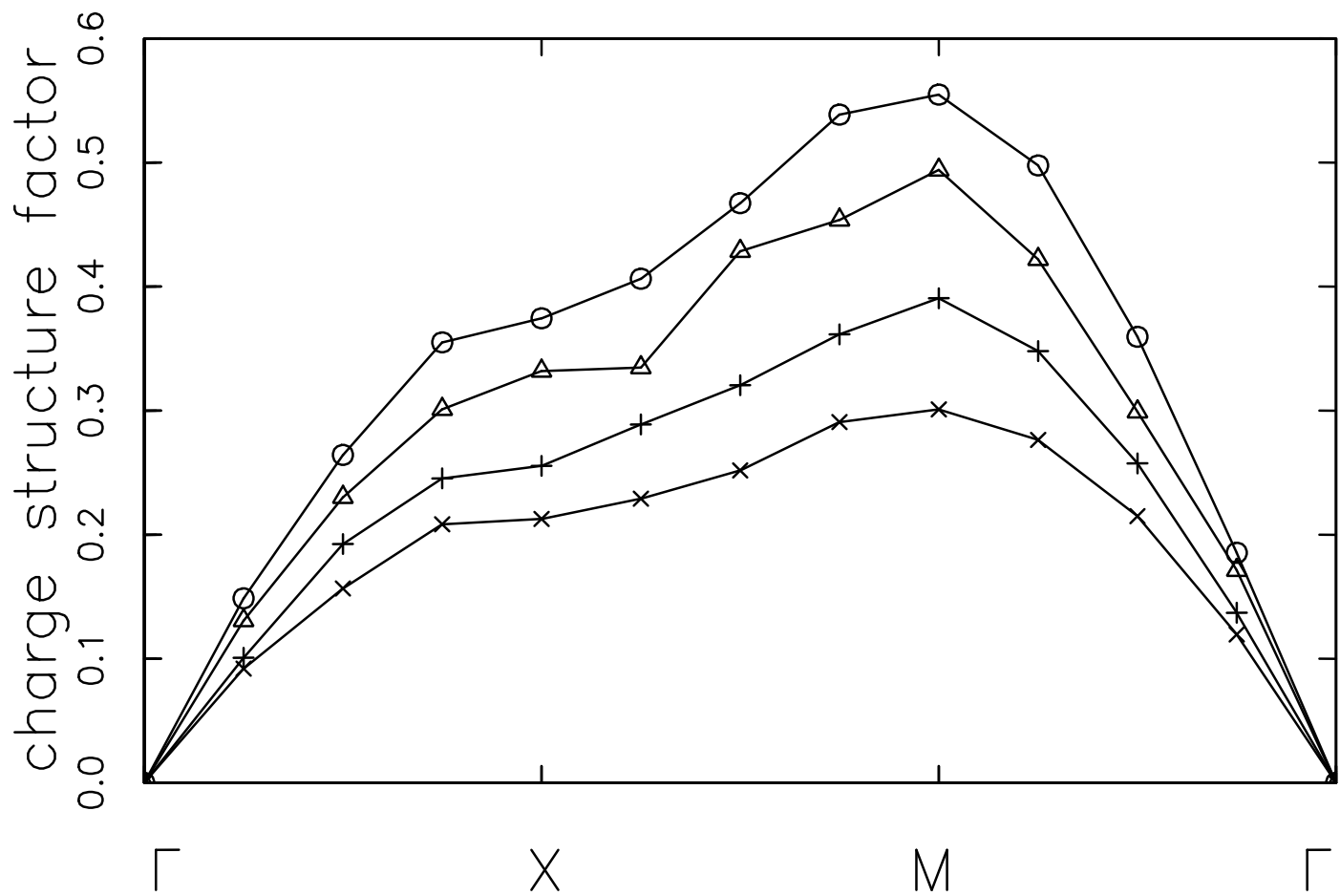
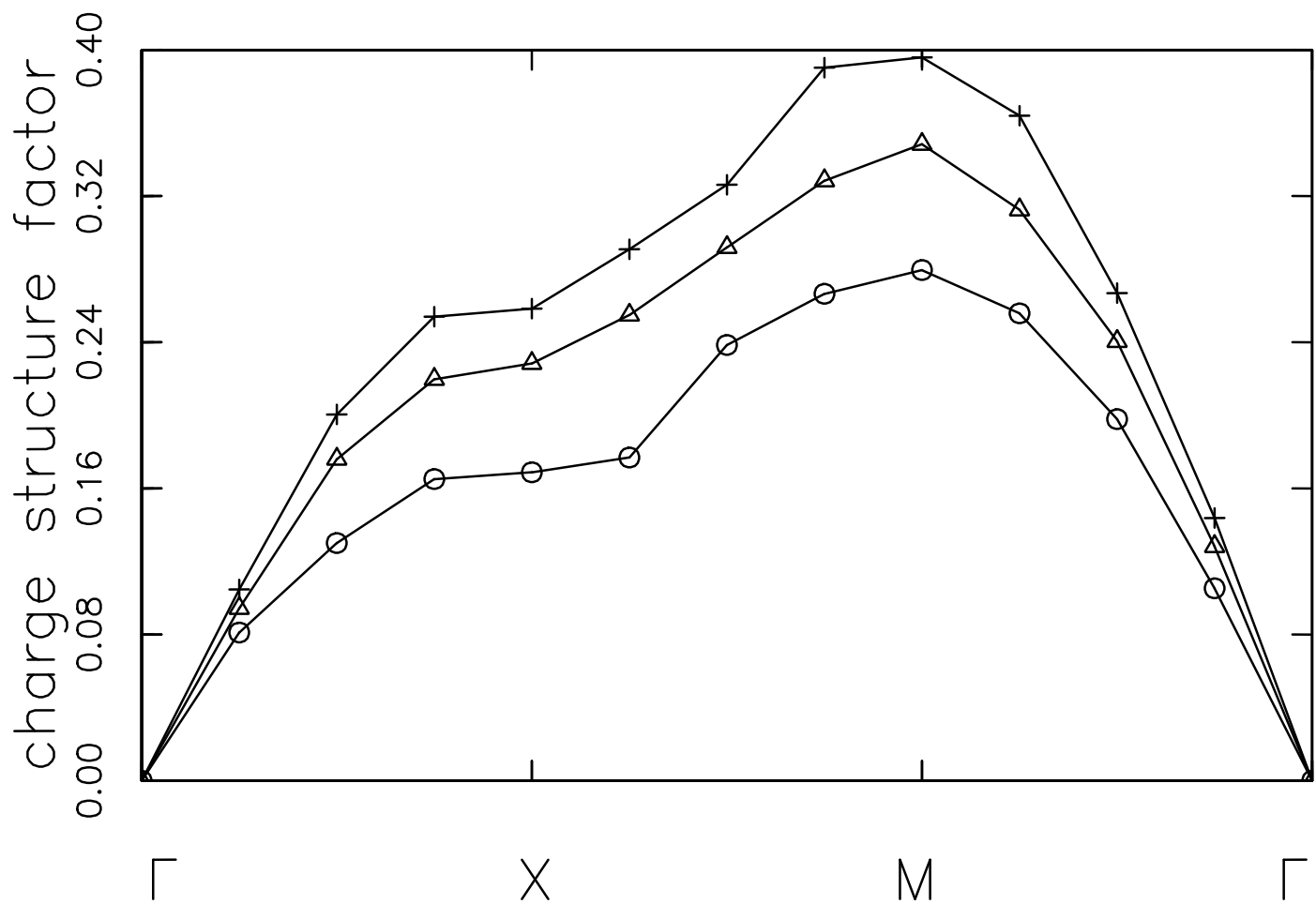
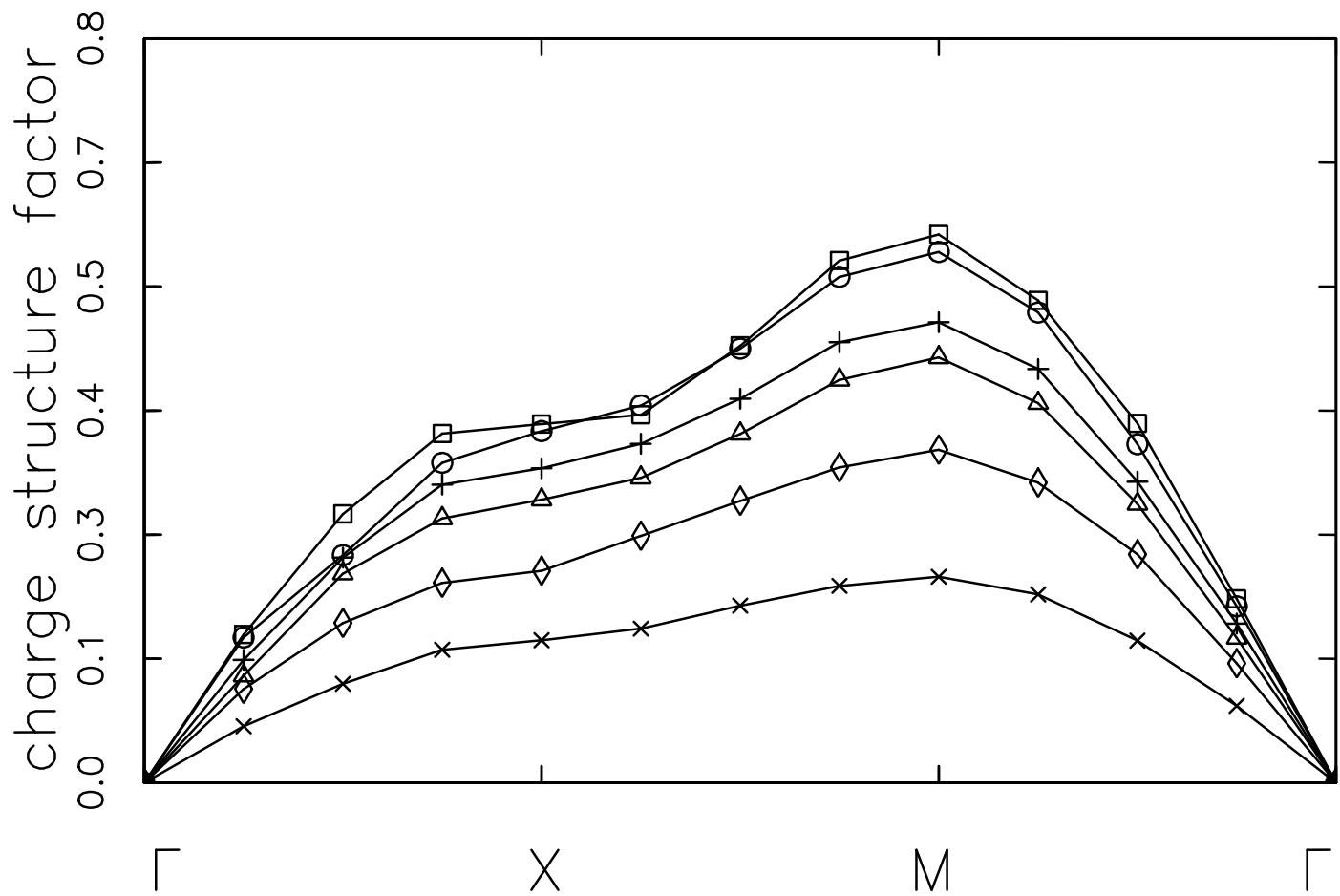
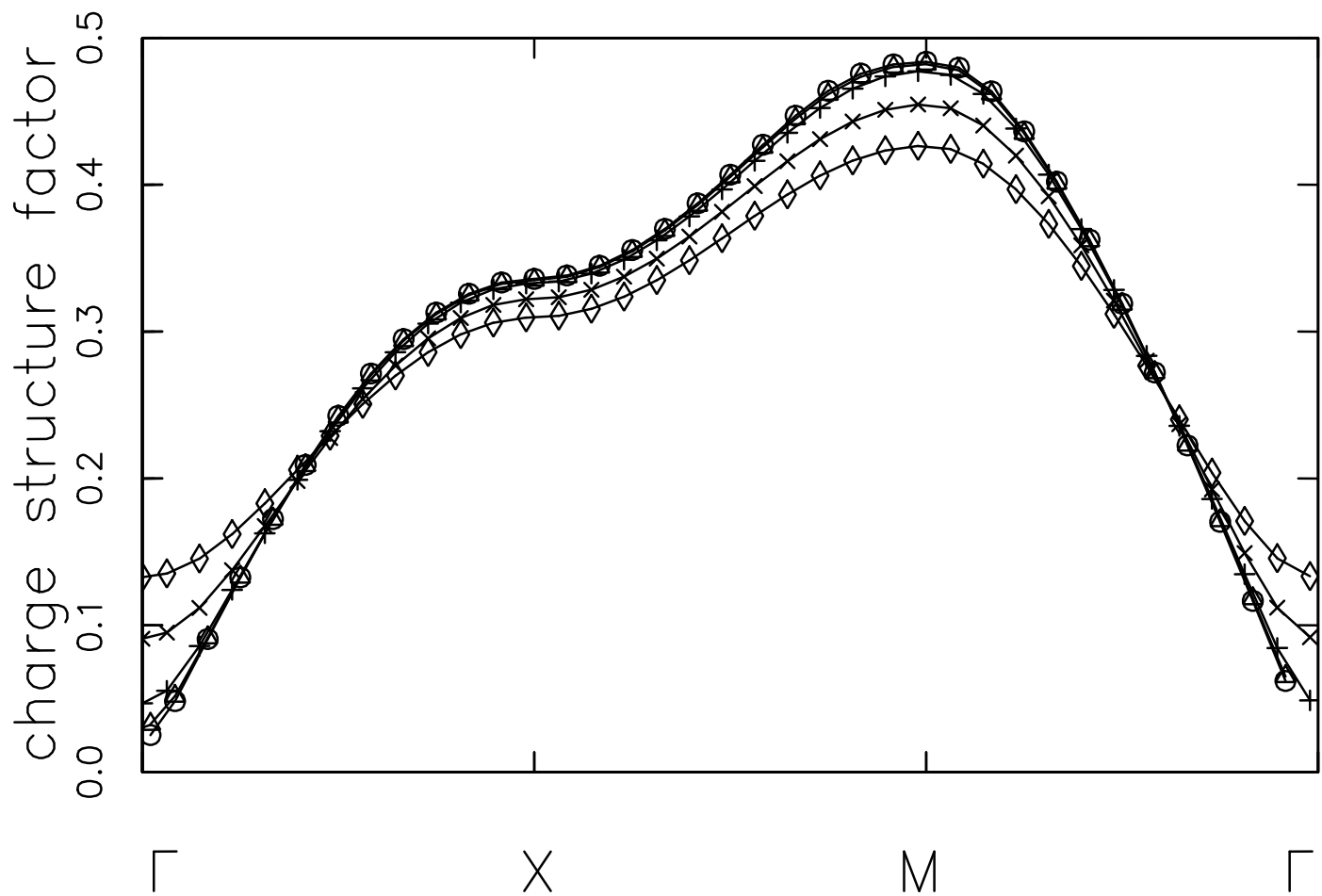
\end{document}